\documentclass[jcp,twocolumn,showpacs,preprintnumbers,superscriptaddress]{revtex4-1}

\usepackage{amsfonts}
\usepackage{graphicx}
\usepackage{amsmath}
\usepackage{amssymb}
\usepackage{latexsym}
\usepackage{psfrag}

\usepackage{dcolumn}
 \usepackage{datetime}
\usepackage{multirow}

\usepackage{subfigure}
\usepackage[bookmarks, colorlinks=true, plainpages = true, linkcolor = blue, citecolor = blue, urlcolor = blue, filecolor = blue]{hyperref}

\begin{document}
\title{Inelastic Tunneling Spectroscopy of Gold-Thiol and Gold-Thiolate Interfaces in Molecular Junctions: The Role of Hydrogen}
 
\author{Firuz Demir} 
\email[Email: ]{fda3@sfu.ca}
\affiliation{Department of Physics, Simon Fraser
University, Burnaby, British Columbia, Canada V5A 1S6}

\author{George Kirczenow} 
\email[Email: ]{kirczeno@sfu.ca}
\altaffiliation{Canadian Institute for Advanced Research, Nanoelectronics Program.}
\affiliation{Department of Physics, Simon Fraser
University, Burnaby, British Columbia, Canada V5A 1S6}

\date{\today}

\begin{abstract}\noindent
It is widely believed that when a molecule with thiol (S-H) end groups bridges a pair of gold electrodes, the S atoms bond to the gold and the thiol H atoms detach from the molecule. However, little is known regarding the details of this process, its time scale, and whether molecules with and without thiol hydrogen atoms can coexist in molecular junctions. Here we explore theoretically how inelastic tunneling spectroscopy (IETS) can shed light on these issues. We present calculations of the geometries, low bias conductances and IETS of propanedithiol and propanedithiolate molecular junctions with gold electrodes. We show that IETS can distinguish between junctions with molecules having no, one or two thiol hydrogen atoms. We find that in most cases the single-molecule junctions in the IETS experiment of Hihath {\em et al.}~[Nano Lett.~\textbf{8}, 1673 (2008)] had no thiol H atoms, but that a molecule with a single thiol H atom may have bridged their junction occasionally. We also consider the evolution of the IETS spectrum as a gold STM tip approaches the intact S--H group at the end of a molecule bound at its other end to a second electrode. We predict the frequency of a vibrational mode of the thiol H atom to increase by a factor $\sim 2$  as the gap between the tip and molecule narrows. Therefore, IETS should be able to track the approach of the tip towards the thiol group of the molecule and detect the detachment of the thiol H atom from the molecule when it occurs.     

\end{abstract}

 
\maketitle

\section{Introduction}
Electrical conduction in molecular junctions consisting of a single organic molecule chemically bonded to a pair of metal contacts has been studied experimentally and theoretically for more than a decade.\cite{review2010} 
However, direct experimental observation of the atomic scale structures of the molecule-electrode interfaces in such devices is not feasible at the present time and the determination of the nature of these interfaces continues to be an important goal in this field.\cite{review2010} The most studied single molecular junctions are those in which the molecule is terminated at its ends by thiol (S--H) groups and bonds to a pair of gold electrodes via the sulfur atoms.\cite{review2010}   
It is widely believed that when such a molecule bonds to gold electrodes, the hydrogen atoms of the thiol groups are displaced by the gold and leave the vicinity of the molecule so that the thiol groups become thiolates.\cite{review2010,Vericat10} However, {\em ab initio} calculations\cite{Gronbeck2000,Ning09,Strange11} have indicated that the energies of such molecular junctions with and without hydrogen atoms attached to the sulfur atoms are very similar, and that junctions with intact S--H groups might even have the lower energies\cite{Ning09,Strange11}, suggesting that some degree of coexistence of species with intact and dissociated S--H groups bound to gold electrodes may be possible.\cite{Gronbeck2000,Ning09,Strange11,Basch04} Furthermore whether the cleavage of the S--H bond occurs on the time scales on which single-molecular junctions are formed (and broken) in statistical STM break junction experiments\cite{expt_stat_STM_2003,Xiao2004,Li2006} 
(an experimental technique widely used today  in studies of single-molecule 
nanoelectronics\cite{review2010}) 
has still not been established.
It is therefore of interest to develop a methodology that can determine unambiguously whether a thiol hydrogen atom is present in the gold-molecule-gold junction at a given time and can provide some information regarding the atomic structure of the thiol-gold region during the process by which the chemical bond between the molecule and the gold electrode forms.
In this paper we explore this issue theoretically and propose inelastic tunneling spectroscopy (IETS) as an attractive candidate for such a methodology.

In single-molecule IETS experiments a bias voltage is applied across a molecular junction at low temperatures and a conductance step is observed in the  current-voltage characteristic of the system when the applied voltage increases beyond the threshold value for the excitation of a vibrational quantum (commonly referred to as a ``phonon") in the junction. 
Many experimental IETS studies  of molecular junctions have been carried out\cite{
Park2000, Kushmerick2004, Qiu2004, WangLeeKretzschmarReed04, Yu2004, Cai2005, Djukic2005, 
Wang2006, Kushmerick2006, Long2006, Naydenov2006, Parks2007, Yu2007, LongTroisi07, Wang2008, deLeon2008,
HihathArroyoRubio-BollingerTao08, KiguchiTalWPKDCvanRuitenbeek08, Song2009, Rahimi2009, Song2009a, Taniguchi2009, Tsutsui2009, Okabayashi2010, Song2010, Arroyo2010, Kim2011, Secker2011, KimarXiv, Kimarstretch, Song2011} and theoretical work\cite{Bonca9597, Emberly2000, May2002, Chen2003, Seideman2003, 
TroisiRatnerNitzan03, Cizek2004, Galperin20040707, Galperin2008, Pecchia2004, 
Pecchia2007, Sergueev2005,TroisiRatner05, Troisi2006, Troisi2006JCP, 
Benesch2006, Jean2006, Paulsson2006, Paulsson2008, Yan2006,
Walczak200607, Frederiksen2007, LaMagna2007, TroisiRatner07, 
Gagliard2007, TroisiBeebe2007, Hartle2008, Kula2008, Luffe2008, 
Jiang2008, Troisi2008, Shimazaki2008, Paulsson.C.F.B.2009, Demir2011, Lin2011bonding,  Lin2011angle, Ueba2007} has accounted for various aspects of the IETS data.
These IETS experiments  together with the theoretical work demonstrated
conclusively that particular molecular species are involved in electrical conduction through metal-molecule-metal junctions. It has also been shown that IETS can be sensitive to the molecular conformation \cite{Rahimi2009, Paulsson.C.F.B.2009, KimarXiv, Demir2012}, the orientation of molecules relative to the electrodes\cite{TroisiRatner07, Lin2011angle} and the gold-sulfur atomic bonding geometries at the molecule-metal interfaces \cite{Demir2011, Lin2011bonding,Demir2012}. 

However, none of the above studies examined the effect that thiol hydrogen atoms, if present, might have on the inelastic tunneling spectra of molecular junctions. We do this in the present article, considering as a specific example molecular junctions with gold electrodes bridged by molecules derived from 1,3-propanedithiol (PDT) that retain one or both thiol hydrogen atoms. We also compare our results for these systems with our previous findings \cite{Demir2011, Demir2012} for 1,3-propanedithiolate molecules (with no thiol hydrogen atoms) bridging gold electrodes and with the experimental  statistical STM break junction IETS data of Hihath \emph{et al.}\cite{HihathArroyoRubio-BollingerTao08}.
We find IETS to be able to distinguish molecules bridging gold electrodes with 
no hydrogen atoms bound to the sulfur atoms of the molecule from molecules with a hydrogen atom bound
to one of the sulfur atoms, and from molecules with hydrogen atoms bound to both sulfur atoms. 
Our results indicate that junctions with no hydrogen atom bound to either sulfur atom of the bridging molecule were the systems probed in most cases by the IETS measurements of Hihath \emph{et al.}\cite{HihathArroyoRubio-BollingerTao08}. We also find the data of Hihath \emph{et al.} to show no evidence of junctions with hydrogen atoms bound to both sulfur atoms of the bridging molecule but to be consistent with transport through molecules with a hydrogen atom bound to {\em only one} of the sulfur atoms being observed occasionally. We also find that a vibrational mode associated with the thiol hydrogen atom has a strong IETS signature and that its frequency is sensitive to the distance between the thiol sulfur atom and the gold electrode to which that sulfur atom bonds. We therefore propose that by tracking the frequency of this vibrational mode  in a statistical STM break junction experiment it should be possible to monitor the closing of the gap between the thiol-terminated end of the molecule and a gold electrode in real time and to detect the detachment of the thiol hydrogen atom from the molecule.

The remainder of this paper is organized as follows: 
In Section~\ref{Theory}, we outline how 
we calculate the relaxed molecular junction geometries, 
vibrational modes, 
low bias elastic tunneling conductances, and 
the conductance step heights associated with inelastic electron scattering in the molecular junctions. 
In Section \ref{Results} we present our results. In Section \ref{relaxation} We discuss the conformations of PDT molecular junctions with different numbers of sulfur atoms having attached hydrogen atoms that we obtained by density functional theory-based unconstrained relaxations of the containing one or two molecules. Our calculated low bias elastic conductances of these junctions are discussed in 
Section~\ref{ETSresults}.
In Section~\ref{IETSresults}, we present the results of our calculations of the inelastic tunneling spectra of these junctions and compare them with the experimental measurements of Hihath \emph{et al.}\cite{HihathArroyoRubio-BollingerTao08} In Section \ref{bondformation} we present our results regarding the evolution of the inelastic tunneling spectrum of a PDT molecule one end of which is bound to a gold electrode while an intact S--H group at the other end is approached by a gold STM tip. We present our conclusions and discuss the experimental implications of our results in Section~\ref{Conclusions}.

\section{Theory}\label{Theory}

The theoretical approach that we adopt here is similar to that used in Refs. \onlinecite{Demir2011, Demir2012}. In the work reported in those papers  \cite{Demir2011, Demir2012} this approach was very successful in accounting for the relevant experimental IETS data \cite{HihathArroyoRubio-BollingerTao08} and identifying for the first time the experimentally realized gold-sulfur bonding geometries in gold-propanedithiolate-gold molecular junctions. An in depth discussion of the underlying theory and the relevant mathematical derivations have been presented in Ref. \onlinecite{Demir2012}. We therefore present only a brief summary of the theoretical approach that we use here and refer the reader to Ref. \onlinecite{Demir2012} for further details.


In this work we focus on inelastic tunneling processes that involve vibrational modes with large amplitudes of vibration of the hydrogen and sulfur atoms that are located at the thiol-gold interface. Therefore, it is necessary to calculate accurate relaxed geometries of both the molecule and nearby atoms of the gold electrodes as well as accurate atomic displacements from equilibrium for these atoms in the vibrational modes. We do this by modeling  the molecular junctions as {\em extended molecules} that consist of the molecule itself and two clusters of gold atoms representing the pair of electrodes to which the molecule binds. Unlike many previous theoretical studies that calculated the relaxed geometries and vibrational modes of only the molecule itself, keeping the positions of the electrode atoms frozen in the geometry of a bulk gold crystal, we calculate the relaxed geometries and vibrational modes of the entire extended molecules including both the molecule and and the gold atoms. The relaxed geometries and vibrational normal modes of these extended molecules were computed using density functional theory\cite{HohenbergKohn64,KohnSham65} as implemented in the GAUSSIAN~09 \cite{Gaussian} package with the B3PW91 density functional\cite{PerdewBurkeErnzerhof1996} and the Lanl2DZ pseudo-potentials and basis set.\cite{Gaussian} Our calculations were repeated for gold clusters of different sizes with up to 14 gold atoms per cluster. These gold cluster sizes were found to be sufficient to achieve convergence of our results for the vibrational mode frequencies and inelastic tunneling intensities with increasing cluster size. A physical reason for this rapid convergence with increasing cluster size is that the frequencies of the vibrational modes that we study are higher than the frequency of the highest phonon mode of bulk gold and for this reason the vibrational modes that we study have a strongly evanescent character in the gold clusters. That is, their vibrational amplitudes in the gold clusters decay very rapidly with increasing distance from the site to which the molecule bonds and are negligible on the furthest gold atoms from the molecule for the larger gold clusters that we consider. In this article we consider fully relaxed extended molecules with no geometrical constraints imposed, and also extended molecules subject to an externally imposed strain that we model by keeping the distance between two outer atoms of the gold clusters fixed during the relaxation of the extended molecule. We note that the experimental inelastic tunneling spectra of Hihath \emph{et al.}\cite{HihathArroyoRubio-BollingerTao08} were taken by sweeping the bias applied across the molecular junction while keeping the STM tip stationary relative to the substrate.

Our calculations of the low bias {\em elastic} conductances of the molecular junctions combine the results of the above {\em ab initio} calculations of the extended molecule geometries with Landauer transport theory,\cite{review2010} Green's function techniques, the solution of the Lippmann-Schwinger equation and the semi-empirical  extended H\"{u}ckel model with the parameters of Ammeter {\em et al.}\cite{yaehmop} to estimate the electronic structures for these geometries. 
As is discussed in detail in Refs.\onlinecite{review2010} and \onlinecite{Cardamone08},
this methodology involves no fitting to any experimental data relating to transport in molecular junctions. It is known to yield low bias conductances in good agreement with experiments for propanedithiolate bridging gold electrodes\cite{Demir2011, Demir2012} as well as other molecules thiol-bonded to gold electrodes \cite{Kushmerick02,Cardamone08,review2010}. We evaluate the zero bias tunneling conductances from the Landauer formula \cite{review2010} 
\begin{equation}\label{tansm } 
{\small 
g = g_0 \sum_{ij}    
\Big\vert t_{ji}^{el} \Big\vert ^2 
\frac{v_j}{v_i} 
}~,
\end{equation}
with the quantum unit of conductance $g_0=2e^2/h$, after evaluating the
elastic transmission amplitudes $t_{ji}^{el}$ from the solution of the Lippmann-Schwinger equation.
In the transmission
 amplitude $t_{ji}$, 
$i$ is the electronic state of a carrier that is coming from the left lead, and
$j$ is the electronic state of a carrier that has been {\em trans}mitted to the right lead. 
$v_j$ and $v_i$ are the corresponding electron velocities.
The coupling of the extended molecule to the electron reservoirs was treated as in previous work
\cite{PivaWolkowKirczenow05, DalgleishKirczenow05, DalgleishKirczenow06, Kirczenow07, PivaWolkowKirczenow08, PivaWolkowKirczenow09, Cardamone08, Cardamone10, Demir2011, Demir2012, Fatemeh11, Fatemeh12}
by attaching a large number of semi-infinite quasi-one-dimensional ideal leads to the valence orbitals of the outer gold atoms of the extended molecule. 

We calculate the IETS intensities (i.e., the conductance step heights) $\delta g_{\alpha}$ associated with the emission of phonons of mode ${\alpha}$ perturbatively, 
in the spirit of an approach proposed by Troisi {\em et al.}\cite{TroisiRatnerNitzan03} who transformed the problem of calculating IETS intensities into an {\em elastic} scattering problem. However, unlike Troisi {\em et al.}\cite{TroisiRatnerNitzan03}, we formulate IETS intensities explicitly in terms of elastic electron transmission amplitudes  $ t_{ji}^{el}$ through the molecular junction. As is shown in Ref. \onlinecite{Demir2012}, we find
\begin{equation} 
\delta g_{\alpha}=  \frac{e^2}{2\pi \omega_{\alpha}} \lim_{A\to0}\sum_{ij} 
\frac{v_j}{v_i }   
\Big\vert \frac{ t_{ji}^{el}(\{A{\bf d}_{{n}\alpha}\})-t_{ji}^{el}
(\{{\bf 0}\})} { A }
 \Big\vert ^2 ~, 
\label{omegaIntensityA} 
\end{equation}
at low temperatures. Eq. \ref{omegaIntensityA} states that the scattering amplitude for {\em inelastic} transmission of an electron through the molecular junction is proportional to the change in the {\em elastic} amplitude for transmission through the junction when its atoms are displaced from their equilibrium positions due to the excitation of vibrational mode $\alpha$.
Here $t_{ji}^{el}(\{{\bf 0}\})$ is the elastic electron transmission amplitude through the molecular junction in its
equilibrium geometry from state $i$ with velocity $v_i$ in the electron source to state $j$ with velocity $v_j$ in the electron drain. ${\bf d}_{{n}\alpha}$ represents the displacements from their equilibrium positions of the atoms $n$ of the extended molecule in normal mode $\alpha$ normalized so that 
$\sum_{n}  m_n  {\bf d}^*_{{n}\alpha'}\cdot {\bf d}_{{n}\alpha} = \delta_{\alpha',\alpha}$ where $m_n$ is the mass of atom $n$. $\omega_{\alpha}$ is the frequency of mode $\alpha$. 
$t_{ji}^{el}(\{A{\bf d}_{{n}\alpha}\})$ is the elastic electron transmission amplitude through the molecular junction with each atom $n$ displaced from its equilibrium position by $A{\bf d}_{{n}\alpha}$, where $A$ is a small parameter.

We evaluate $ t_{ji}^{el}$ in Eq.~(\ref{omegaIntensityA}) numerically to find the heights $\delta g_{\alpha}$~of the conductance steps that arise from inelastic tunneling processes due to emission of phonons of vibrational mode $\alpha$.
The calculations are carried out at the zero bias Fermi energy since the values of the bias at which the inelastic transmission occurs in the experiments of Hihath \emph{et al.} \cite{HihathArroyoRubio-BollingerTao08} are low.

The validity of our methodology rests largely on the fact
that we rely on {\em ab initio} density functional theory to calculate 
the frequencies and atomic displacements from equilibrium  for the vibrational 
normal modes that give rise to the features of interest in the inelastic tunneling spectra. 
Such calculations are accurate because in the Born-Oppenheimer
approximation they constitute electronic {\em ground state total energy}
calculations for which density functional theory is appropriate\cite{review2010, HohenbergKohn64, KohnSham65} 
and for which the density functional model that we use\cite{PerdewBurkeErnzerhof1996, Gaussian} has been optimized.
Our key predictions concern these
accurately calculated frequencies (or the associated phonon energies) for the modes 
with the largest intensities in the inelastic tunneling spectra, and the dependence 
of these frequencies on the details of the conformation of the
molecular junction. We also calculate 
the junction conformations 
using density functional theory. We rely on the extended H\"{u}ckel model
and perturbation theory only to determine which of the vibrational modes
have the largest inelastic tunneling intensities. It should also be noted
that perturbative calculations of this type are appropriate provided that the transmitted
electrons have energies far from any elastic resonance and if the amplitudes
of the excited vibrational modes are small. These conditions are expected to be
satisfied at low bias voltages and cryogenic temperatures such as those
in the experiment of Hihath \emph{et al.} \cite{HihathArroyoRubio-BollingerTao08}
Also, the extended H\"{u}ckel model that we use has been found to yield
results in reasonably good agreement with experiment for 
elastic\cite{review2010, Demir2011, Demir2012, Kushmerick02, Cardamone08,Emberly2001,Emberly2001a} 
and inelastic\cite{Demir2011, Demir2012}
transport in a variety of molecular wires thiol-bonded to gold electrodes.

\section{Results}
\label{Results}
\subsection{Conformations of extended molecules obtained by unconstrained relaxations}
\label{relaxation}

We considered extended molecules where the molecule bonds via its sulfur atoms to finite clusters of gold atoms, and relaxed these entire structures using density functional theory-based calculations.\cite{HohenbergKohn64, KohnSham65, PerdewBurkeErnzerhof1996,  Gaussian} 
We carried out the density functional geometry relaxations starting from initial trial geometries of  the PDT molecule with a hydrogen atom attached to one, both or neither of the two sulfur atoms of the molecule.
\begin{figure}[t]
\centering
\includegraphics[width=1.0\linewidth]{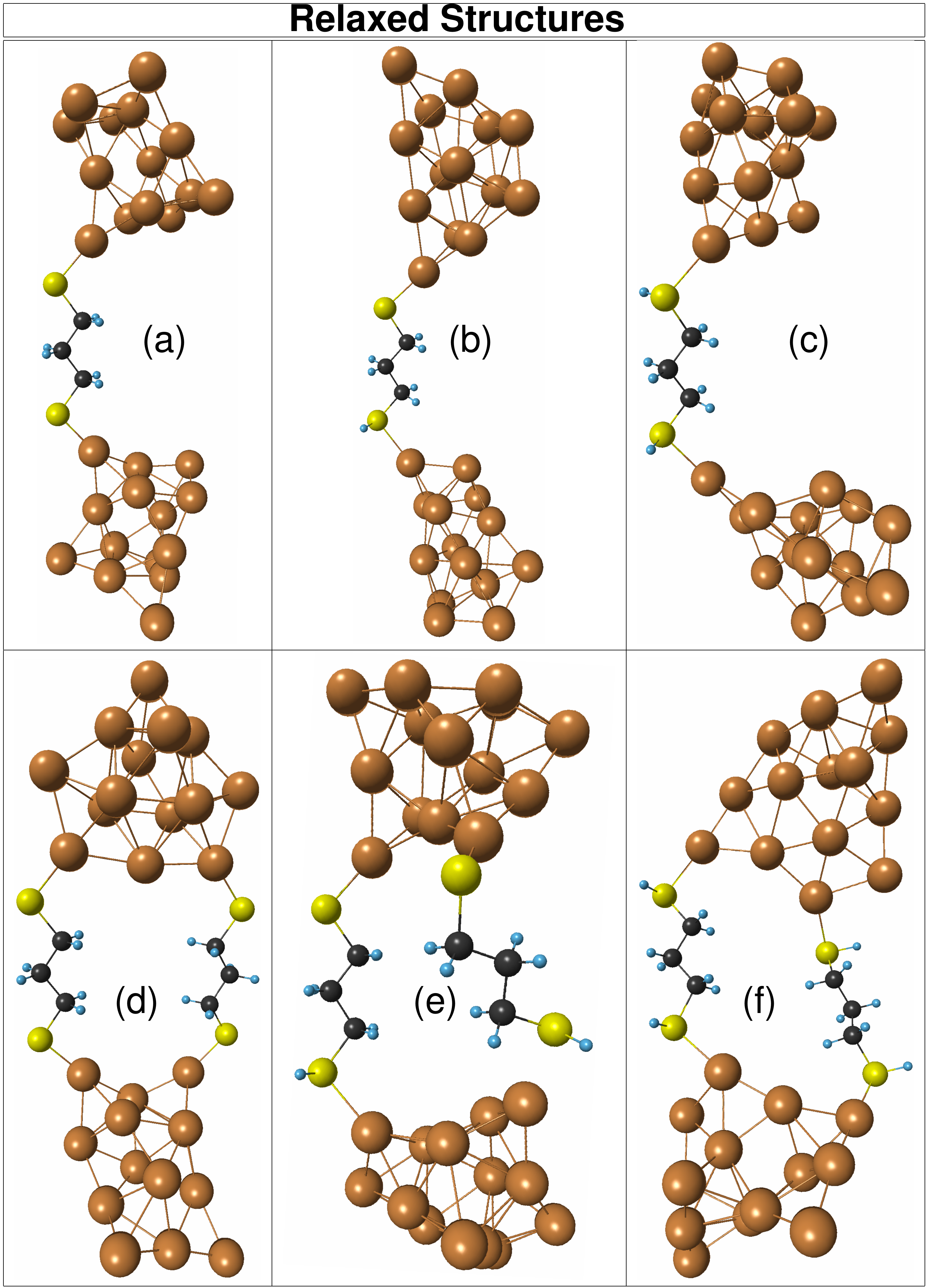}
\caption{{\color{red}{(Color online.) }} Examples of relaxed structures of trans-PDT molecules bonded to gold clusters in the top geometry.\cite{Macmolplt}
(a): A PDT molecule with no hydrogen atom attached to either sulfur atom.  
(b): A PDT molecule with a hydrogen atom attached to one of the sulfur atoms.
(c): A PDT molecule with a hydrogen atom attached to each of the two sulfur atoms.
(d): Two PDT molecules bridging the gold clusters with no hydrogen atom attached to any sulfur atom. 
(e): Two PDT molecules bridging the gold clusters with a hydrogen atom attached to one of the sulfur atoms
of each molecule. The sulfur atom of the molecule on the right has not chemisorbed to the lower gold cluster while each of the other three sulfur atoms in the system has chemisorbed to a gold cluster. 
(f): Two PDT molecules bridging the gold clusters with a hydrogen atom attached to each of the two sulfur atoms
of each molecule.
Carbon, hydrogen, sulfur and gold atoms are black, blue,
yellow and amber, respectively.
}
\label{Fig1}
\end{figure}
%
\begin{table} 
\centering
\caption{{\color{red}{(Color online.) }}
Calculated low bias elastic conductances, and mode I phonon energies and IETS intensities for
the relaxed structures of one and two molecules with different numbers of thiol hydrogen atoms
bridging gold atomic clusters. The numerical values labelled (a) -- (f) are for the molecular junctions shown in
Fig. \ref{Fig1} (a) -- (f), respectively. The experimentally measured low bias conductance values of Hihath \emph{et al.} for single-molecule PDT junctions were $\sim 0.006 g_{0}$ $\pm 0.002$.     \cite{HihathArroyoRubio-BollingerTao08} $g_0 = 2e^2/h$. 
}
\label{CalculatedConductance}
\includegraphics[width=1.0\linewidth]{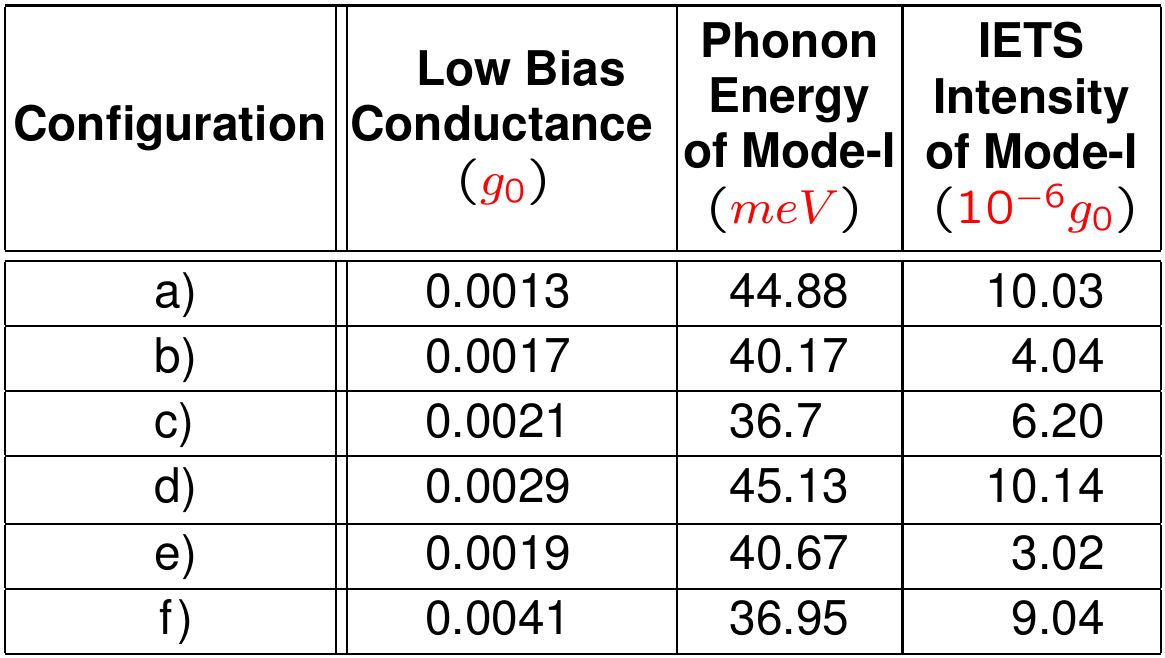}
\end{table}

We were able to obtain relaxed structures in which the hydrogen atom(s) remaining attached to one or both of the sulfur atom(s) for bonding geometries in which the sulfur with the attached hydrogen atom bonds to one gold atom (referred to as ``top site'' bonding geometries) but not for geometries in which the sulfur bonds to two or three gold atoms. Therefore in the remainder of this paper we will consider only top site bonded geometries; the two other types of bonding geometries for PDT molecules with no attached thiol hydrogen atoms have been discussed in Refs \onlinecite{Demir2011, Demir2012}. 

Representative examples of relaxed PDT extended molecules with the trans conformation are shown in Fig.~\ref{Fig1}. In Fig.~\ref{Fig1}, all S atoms bond to the gold clusters in the top site geometry. 
In all of the examples shown in Fig.~\ref{Fig1} each gold cluster contains 14 atoms and the
relaxations were performed with no constraints applied to the extended molecule geometries.
In Fig.~\ref{Fig1} (a), (b) and (c) a single PDT molecule with zero, one and two thiol hydrogen atoms bridges the gold clusters. In Fig.~\ref{Fig1} (d), (e) and (f) two PDT molecules, each with zero, one and two thiol hydrogen atoms, bridge the gold clusters.

Our findings regarding the energetics of the structures shown in Fig.~\ref{Fig1} are qualitatively similar to those of the previous studies\cite{Gronbeck2000,Ning09,Strange11,Basch04} that have indicated that the energies of molecular junctions with and without hydrogen atoms attached to the sulfur atoms are very similar, and that junctions with intact S--H groups might even have the lower energies. For example, our density functional theory-based calculations yielded an energy for the structure  in Fig.~\ref{Fig1} (a) (that has no thiol H atoms) that, together with the energy
of a free H${_2}$ molecule, is higher than the energy of the structure  in Fig.~\ref{Fig1} (c) (that has two
thiol H atoms) by 0.27eV. Similarly, we found an energy for the structure  in Fig.~\ref{Fig1} (d) (that has no thiol H atoms) that, together with the energy
of two free H${_2}$ molecules, is higher than that of the structure  in Fig.~\ref{Fig1} (f) (that has four
thiol H atoms) by 0.04eV. This suggests that a system with a dithiol molecule bridging gold electrodes may be slightly more stable energetically than that formed if the molecule dissociates into the dithiolate and a free H${_2}$ molecule. However, this ignores entropy considerations, which favor the latter system and furthermore it
is not known whether the H atoms that detach from the thiol groups form  H${_2}$ molecules or other, possibly more stable, moieties. It is also worth noting that the geometries of the gold clusters in  Fig.~\ref{Fig1} (a) and (c) are not identical, as is also the case for Fig.~\ref{Fig1} (d) and (f), and that conclusions drawn from the above comparisons between the energies of these structures should be approached with some caution for this reason as well.

\subsection{Elastic conductances in the limit of low bias}  \label{ETSresults}
Our calculated zero bias conductance values for the molecular junctions shown in Fig.~\ref{Fig1} are
presented in in Table~\ref{CalculatedConductance} where rows (a) -- (f) correspond to structures
(a) -- (f) of Fig.~\ref{Fig1}, respectively.

As is seen in Table~\ref{CalculatedConductance}, 
the calculated zero bias conductance values  are 
g=0.0013$g_0$, 0.0017$g_0$ and  0.0021$g_0$ for top-site bonded single-molecule junctions with no thiol hydrogen atoms, one thiol hydrogen atom, and two thiol hydrogen atoms, respectively. 
All of these theoretical values agree with the experimental
value $0.006 \pm 0.002 g_0$ of Hihath {\em et al.}\cite{HihathArroyoRubio-BollingerTao08} to a degree that is
typical of the experimental and theoretical literature\cite{Lindsay07} for molecules thiol
bonded to gold electrodes. These calculated elastic conductance values show an
increasing trend with increasing number of thiol hydrogen atoms in the molecular junction.
However, since the calculated values differ by very modest amounts (less than a factor of two), 
determining experimentally whether or not hydrogen atoms remain bound to the sulfur atom(s) in experimental realizations of gold-PDT-gold molecular junctions on the basis of low bias conductance measurements alone is expected to be difficult or impossible.   
As shown in Section \ref{IETSresults}, comparing our predictions for the {\em inelastic} tunneling spectra with experiment\cite{HihathArroyoRubio-BollingerTao08} provides strong evidence that PDT molecular junctions with no thiol hydrogen atoms predominated in the experiments of Hihath {\em et al.}\cite{HihathArroyoRubio-BollingerTao08}.

In the majority of cases the calculated zero bias conductance values in Table~\ref{CalculatedConductance} for molecular junctions with pairs of molecules bridging the gold contacts in parallel are roughly a factor of 2 larger than the conductances of the corresponding single molecule junctions, as might be expected for pairs of molecules\cite{Yaliraki98, Magoga99, Lang2000, Lagerqvist2004, Liu2005, Geng2005, Long2007,
Wang2010, Reuter2011, Demir2012} or other quantum conductors\cite{Castano90} connecting electron reservoirs. An exception is the structure in  Fig.~\ref{Fig1}(e) for which the calculated low bias conductance is close to that of the corresponding single-molecule junction in Fig.~\ref{Fig1}(b). We attribute this different behavior to the molecule on the right in Fig.~\ref{Fig1}(e) not having formed a chemical bond between the sulfur atom with the attached hydrogen atom and the nearby gold cluster, resulting in poor conduction through that molecule.

\subsection{Inelastic Tunneling Spectroscopy of Thiol Groups in Unconstrained Molecular Junctions}
\label{IETSresults}
We will begin by considering the IETS of extended molecules that have been relaxed without any geometrical constraints being imposed during the relaxation process as discussed in Section \ref{relaxation}.
Since our objective is to understand how the presence of thiol hydrogen atoms should affect the inelastic tunneling spectra of molecular junctions we will focus on the IETS signatures of those vibrational modes that have the strongest amplitudes of vibration on S atoms or thiol H atoms and the strongest IETS intensities associated with those modes. We find the relevant modes to fall within the energy range from $25$ to $60$~meV.
\unitlength=0.01\linewidth
\begin{figure}[ht!] 
\centering
\subfigure[] {\label{Fig2(a)} 
\includegraphics[width=0.95\linewidth]{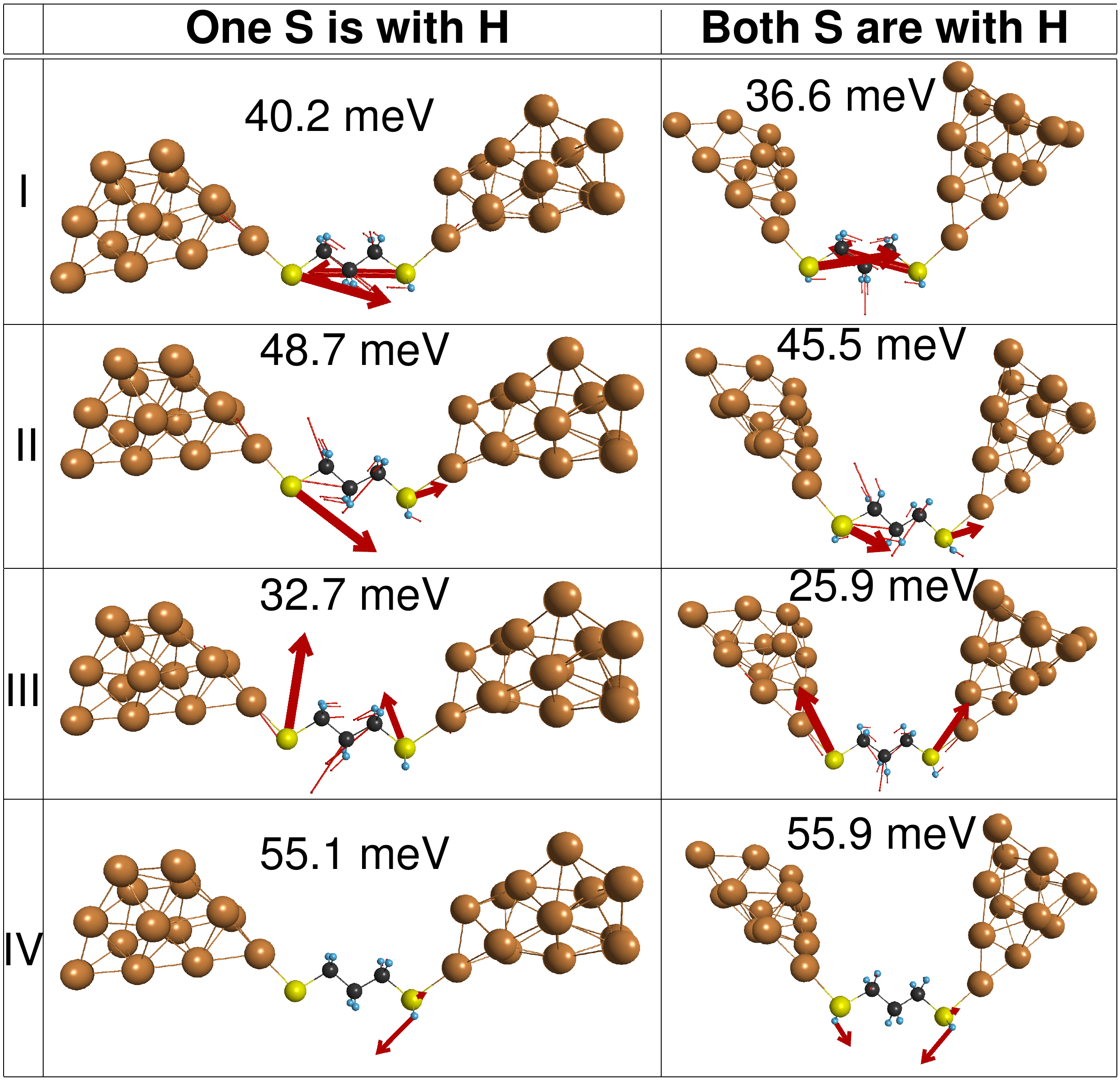}     }
\subfigure[] {\label{Fig2(b)}
\includegraphics[width=0.95\linewidth]{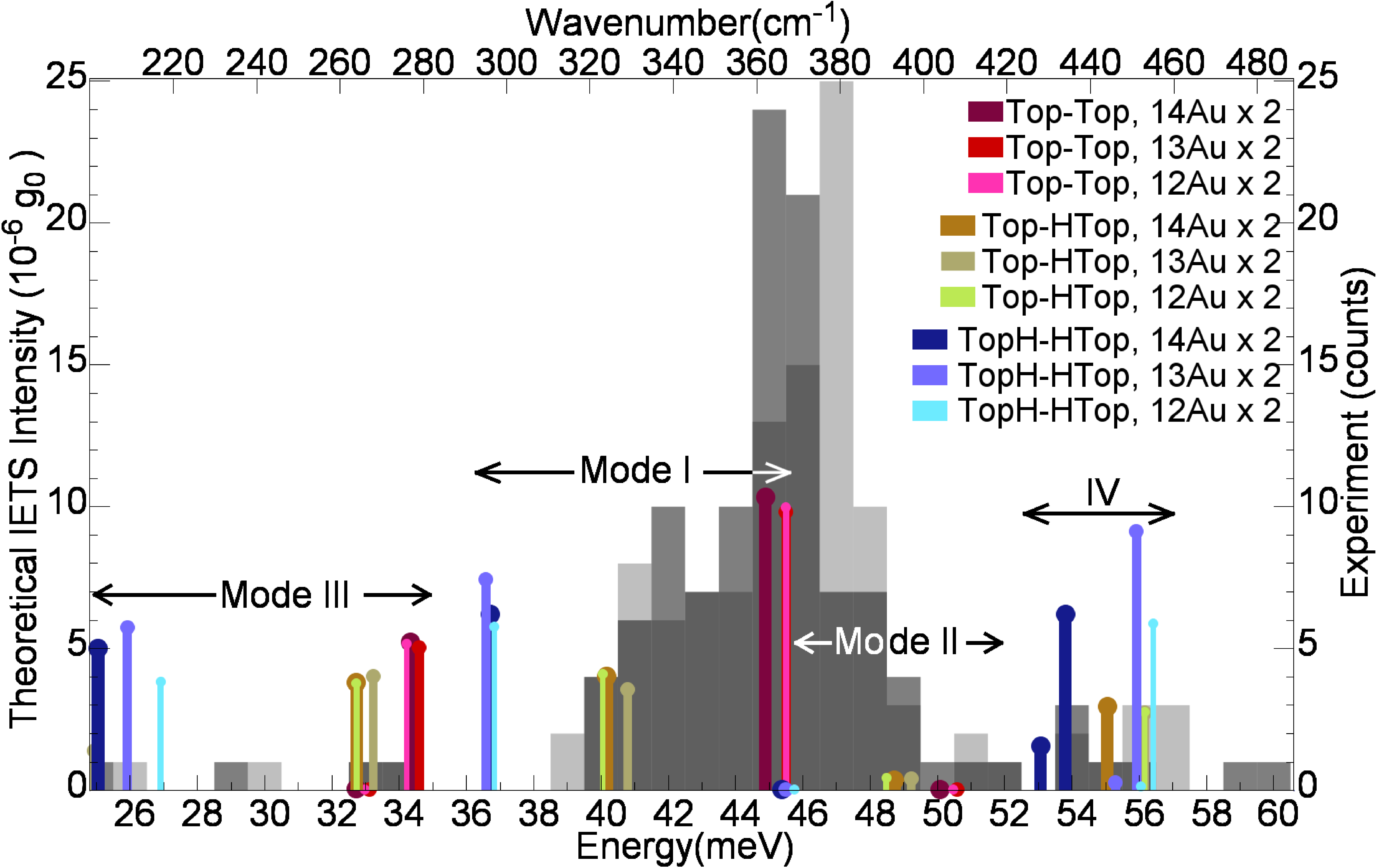}     }
\caption{{\color{red}{(Color online.) }}
\subref{Fig2(a)} 
Calculated vibrational modes in the energy range from $25$ to $60$~meV for PDT bridging gold nano-clusters with hydrogen atoms bound to one or both sulfur atom(s), the sulfur atoms being bonded to gold in top-site geometries. These structures were relaxed with no constraints imposed on their geometries.
Carbon, hydrogen, sulfur and gold atoms are black, blue, yellow and amber, respectively.\cite{Macmolplt} 
Arrows show un-normalized atomic displacements; the heavier arrows indicate the motion of the sulfur atoms. 
\subref{Fig2(b)} 
Calculated IETS intensities (colored) vs. calculated vibrational mode energies for PDT molecules linking pairs of gold clusters with between 12 and 14 Au atoms in each cluster. 
Results are shown for 
both sulfur atoms bonding to the gold in top-site geometries and having no hydrogen atom attached to either sulfur atom, having a hydrogen atom attached to one sulfur atom only,  and 
having a hydrogen atom attached to each of the sulfur atoms.  
The experimental IETS phonon mode histogram of Hihath \emph{et al.}\cite {HihathArroyoRubio-BollingerTao08} is shown in (darker, lighter) grey for (positive, negative) bias voltages.
The energy ranges in which modes of types I, II, III, and IV in \subref{Fig2(a)} occur are indicated by arrows. 
}
\end{figure}

Representative examples of these vibrational modes and their energies are shown in Fig.~\ref{Fig2(a)} for PDT sulfur atoms bound to gold clusters in top site conformations (as discussed in Section \ref{relaxation}) with a hydrogen atom bound to one or both sulfur atoms. The modes are classified according to the nature their most prominent atomic vibrations. In modes I, II and III the largest vibrational amplitudes are those of the sulfur atoms.
In mode I and II the two sulfur atoms move in antiphase and in phase, respectively, approximately along the axis of the molecule. (These modes are sometimes referred to as the symmetric and antisymmetric stretch modes, respectively.) In mode III the sulfur atoms move in phase, in directions not aligned with the molecular axis. In mode IV the most prominent atomic vibrations are those of the thiol hydrogen atoms whose motion is roughly torsional relative to the direction of the nearest sulfur-carbon bond.

The calculated IETS intensities and energies for these modes are shown in Fig.~\ref{Fig2(b)} together with 
the experimental IETS histogram of Hihath \emph{et al.} \cite {HihathArroyoRubio-BollingerTao08}.  
Our corresponding results for top site bonded PDT wires with no hydrogen atoms attached to the sulfur\cite{Demir2011, Demir2012} are also shown in Fig.~\ref{Fig2(b)} for comparison. Notice that the calculated IETS intensities corresponding to mode II in Fig.~\ref{Fig2(b)} are much smaller than those for the other modes, due to an approximate symmetry that results in a partial cancellation between the terms $t_{ji}^{el}$ in Eq.~(\ref{omegaIntensityA}) for mode II, as has been discussed in Refs \onlinecite{Demir2011, Demir2012, Troisi2006JCP} for the case when no thiol hydrogen atoms are present.   

As can be seen in Fig.~\ref{Fig2(b)}, the effect of attaching H atoms to the molecular S atoms on the frequency of vibrational mode I (the high intensity mode that has been shown \cite{Demir2011, Demir2012} to play the dominant role in the experiment of Hihath \emph{et al.} \cite {HihathArroyoRubio-BollingerTao08})  is quite dramatic: For the top site bound PDT junction with {\em no} H atom bound to either S atom, this mode occurs at 45.5~meV 
in Fig.~\ref{Fig2(b)} and is responsible for the strongest peak visible in the experimental histogram\cite {HihathArroyoRubio-BollingerTao08}.  However, for the three examples of molecular wires each with {\em one hydrogen atom attached to one sulfur atom only}, mode I can be seen around 
40~meV 
in Fig.~\ref{Fig2(b)}, where some counts were recorded in the experiment\cite {HihathArroyoRubio-BollingerTao08}. 
The calculated IETS intensity for mode I for these structures is somewhat weaker than that for the junctions with no thiol hydrogen atoms and is very similar to that of mode III near 33 meV where very few counts were recorded in the experiment. This suggests while a small fraction of the PDT molecular junctions in the experiment\cite {HihathArroyoRubio-BollingerTao08} may have had a hydrogen atom attached to one of the sulfur atoms, most of the counts near 40 meV in the experimental histogram\cite {HihathArroyoRubio-BollingerTao08} were probably {\em not} due to molecules with a hydrogen atom attached to one of the sulfur atoms. That is, PDT molecules with no thiol hydrogen atoms and with the sulfur atoms each bonding to two atoms of the gold electrodes (the pure bridge geometry) are the likely explanation of most of the experimental counts near 40 meV since these structures have {\em strong} calculated IETS intensities in that energy range as has been
discussed in Ref.\onlinecite{Demir2012}. 

Notice that no counts are seen in the experimental histogram\cite {HihathArroyoRubio-BollingerTao08} 
(Fig.~\ref{Fig2(b)}) for phonon energies near 
36.7~meV 
in Fig.~\ref{Fig2(b)}, where  we calculate mode I  for the top site bound PDT junction with a H atom bound to {\em each} of the S atoms. This indicates that molecules with two thiol hydrogen atoms did not contribute significantly to the observed inelastic tunneling spectra of Hihath \emph{et al.} \cite {HihathArroyoRubio-BollingerTao08}.

The top site bonded molecular wires with a H atom bound to one or both of their S atoms all exhibit a vibrational mode (mode IV in Fig.~\ref{Fig2(a)}) that involves primarily the motion of those hydrogen atoms. We find this mode to be located in the range 
53~-~
56~meV.~
The experimental histogram\cite {HihathArroyoRubio-BollingerTao08} shows a few counts in this range and its vicinity. 
Thus the weak feature in the experimental histogram in this range might possibly be due to a small minority of the molecular junctions in the experiment of Hihath {\em et al.}\cite {HihathArroyoRubio-BollingerTao08} having a hydrogen atom attached to a sulfur atom, although gauche conformations of the PDT molecule with no thiol hydrogen atoms may also have contributed to the observed inelastic tunneling spectra in that energy range, as has been discussed in Ref.\onlinecite{Demir2012}.  Further IETS experiments with greater sensitivity to weak features in the inelastic spectra will be required to discriminate definitively between these possibilities.

In the preceding discussion we have focussed on cases where a single molecule connects the two electrodes. However in statistical STM break junction experiments such as that of Hihath \emph{et al.}\cite{HihathArroyoRubio-BollingerTao08} the electrodes can also be bridged simultaneously by two (or more) molecules  in parallel.  Here we report theoretical results for the inelastic contributions to the conductances of some examples of systems of this kind. 
%
\unitlength=0.01\linewidth
\begin{figure}[t] 
\centering
\includegraphics[width=1.01\linewidth]{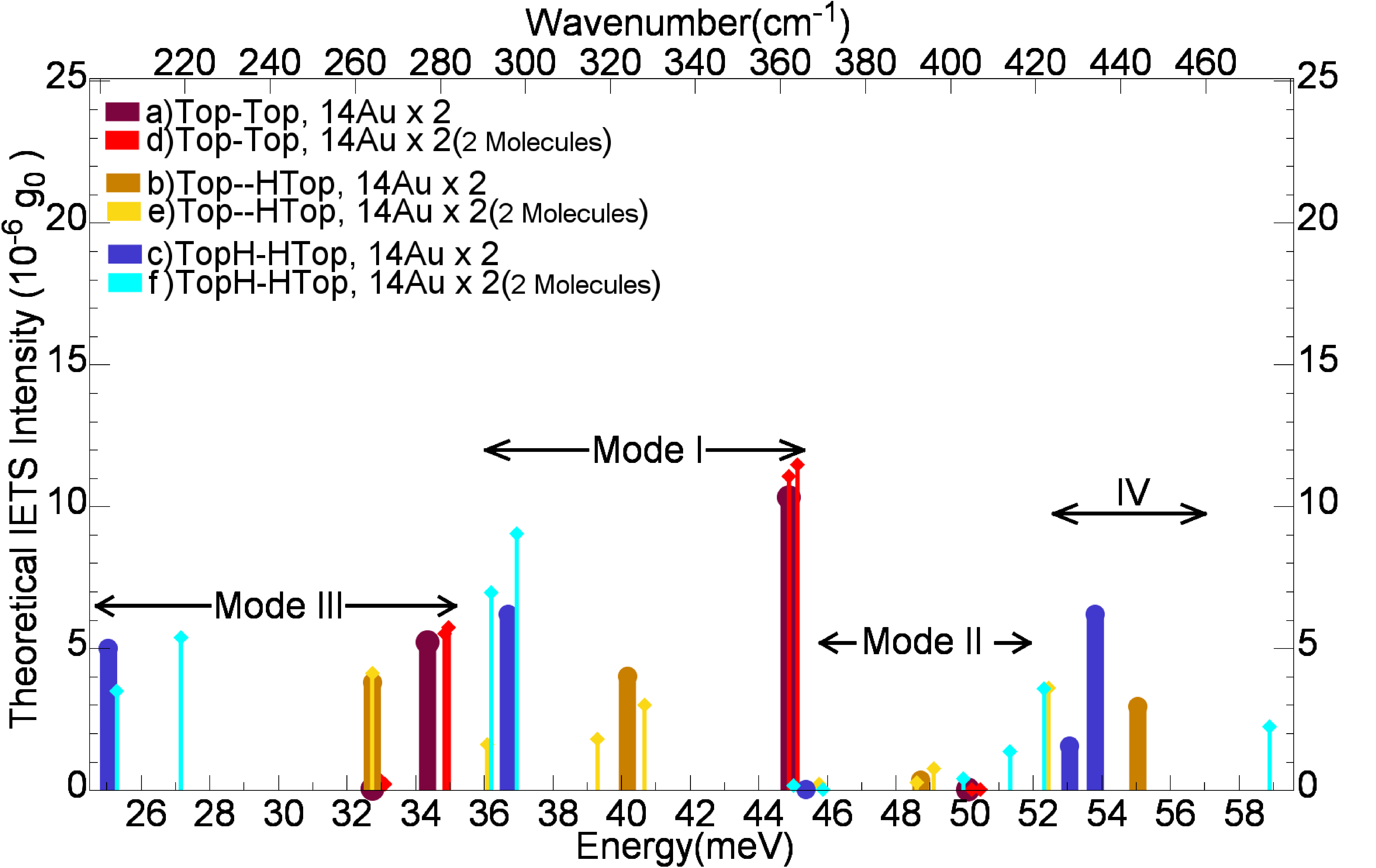}   
\caption{{\color{red}{(Color online.) }}
Calculated inelastic tunneling spectra for one and two PDT molecules
with different numbers of thiol hydrogen atoms
bridging pairs of 14 atom gold clusters in the fully relaxed geometries
depicted in Fig. \ref{Fig1}. Labels (a) -- (f) denote the structures (a) -- (f)
of Fig. \ref{Fig1}. The atomic motions in modes I -- IV are similar to those
shown for modes I -- IV of the single molecules in Fig. 2(a).
}
\label{Fig3(b)}
\end{figure}
%
We show 
a pair of top site bonded molecules having no hydrogen atom attached to either sulfur atom 
in Fig~\ref{Fig1}(d), 
a pair of top site bonded molecules having a hydrogen atom attached to one sulfur atom of each molecule in Fig~\ref{Fig1}(e), and 
in Fig~\ref{Fig1}(f) a pair of top site bonded molecules having a hydrogen atom attached to each sulfur atom of each molecule. 
The intermolecular distances defined as the shortest distances between hydrogen atoms belonging to the backbones of the two molecules are 4.48, 3.21, and 2.28 Angstroms respectively. The calculated inelastic 
tunneling spectra for these structures are shown together with those of the corresponding single molecule junctions
( Fig~\ref{Fig1}(a),(b) and (c) respectively) in Fig.~\ref{Fig3(b)}. As can be seen in Fig.~\ref{Fig3(b)},
the calculated inelastic tunneling spectra for these pairs of molecular wires are very similar to the spectra of the individual wires apart from small frequency shifts and splittings due to the weak interactions between the molecules.
It is worth noting that although one of the molecules in Fig~\ref{Fig1}(e) is poorly bonded to one of the gold
electrodes and contributes little to the low bias elastic conductance of the system (as has been discussed in Section 
\ref{ETSresults}) there is still a significant splitting of the mode I frequency for this two-molecule junction in  Fig.~\ref{Fig3(b)} due to the coupling between the vibrational modes of the two molecules. 

\subsection{Inelastic Tunneling Spectroscopy of Bond Formation Between a Gold Electrode and a Thiol-Terminated Molecule}
\label{bondformation}

The preceding results have been for extended molecules that were relaxed with no geometrical constraints. Thus they can be regarded as pertaining to molecular junctions that have reached a local energy minimum, having formed chemical bonds between the molecule(s) and the gold electrodes. In this Section we will consider a different situation, namely, molecular junctions in the process of bond formation. More specifically, we will examine the evolution of the inelastic tunneling spectrum as a PDT molecule that has bonded to one gold electrode via a sulfur atom is approached by second gold electrode (that may be an STM tip) so that chemical bonding between the molecule and the second electrode is in the process of taking place. 
%
\unitlength=0.01\linewidth
\begin{figure*}[ht!] 
\centering
\subfigure[] {   \label{Fig4(a)} 
\includegraphics[width=1.00\linewidth]{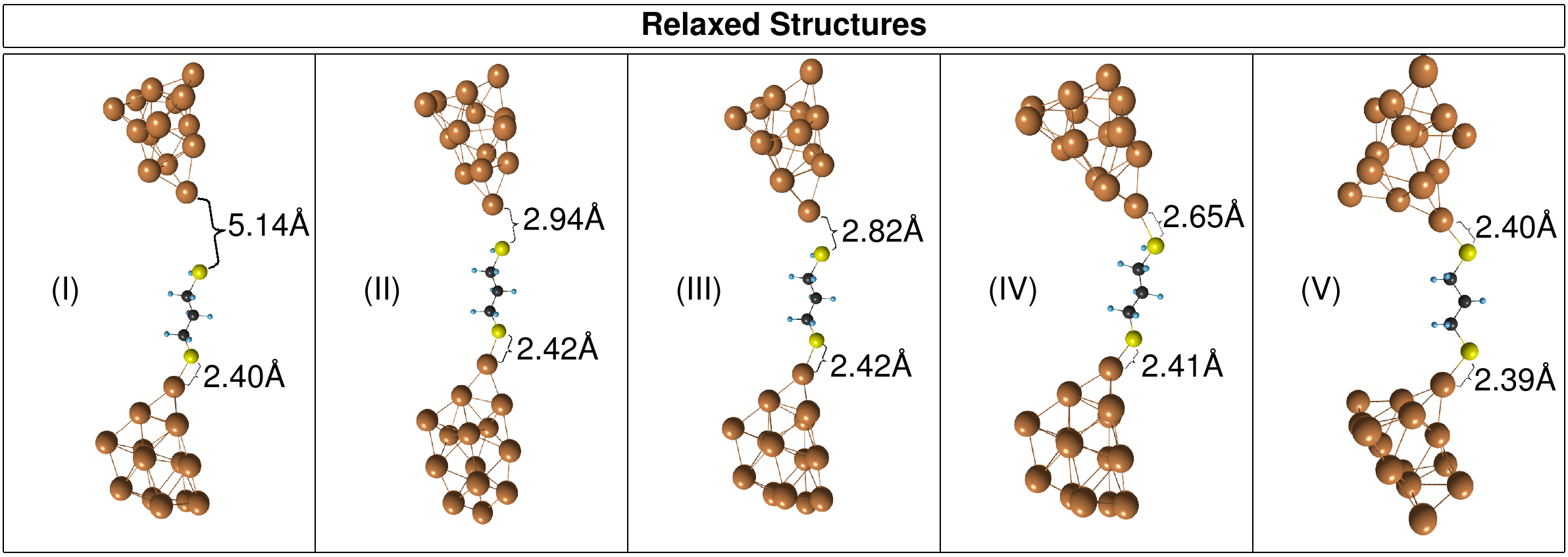}     }
\subfigure[] {   \label{Fig4(b)}
\includegraphics[width=0.485\linewidth]{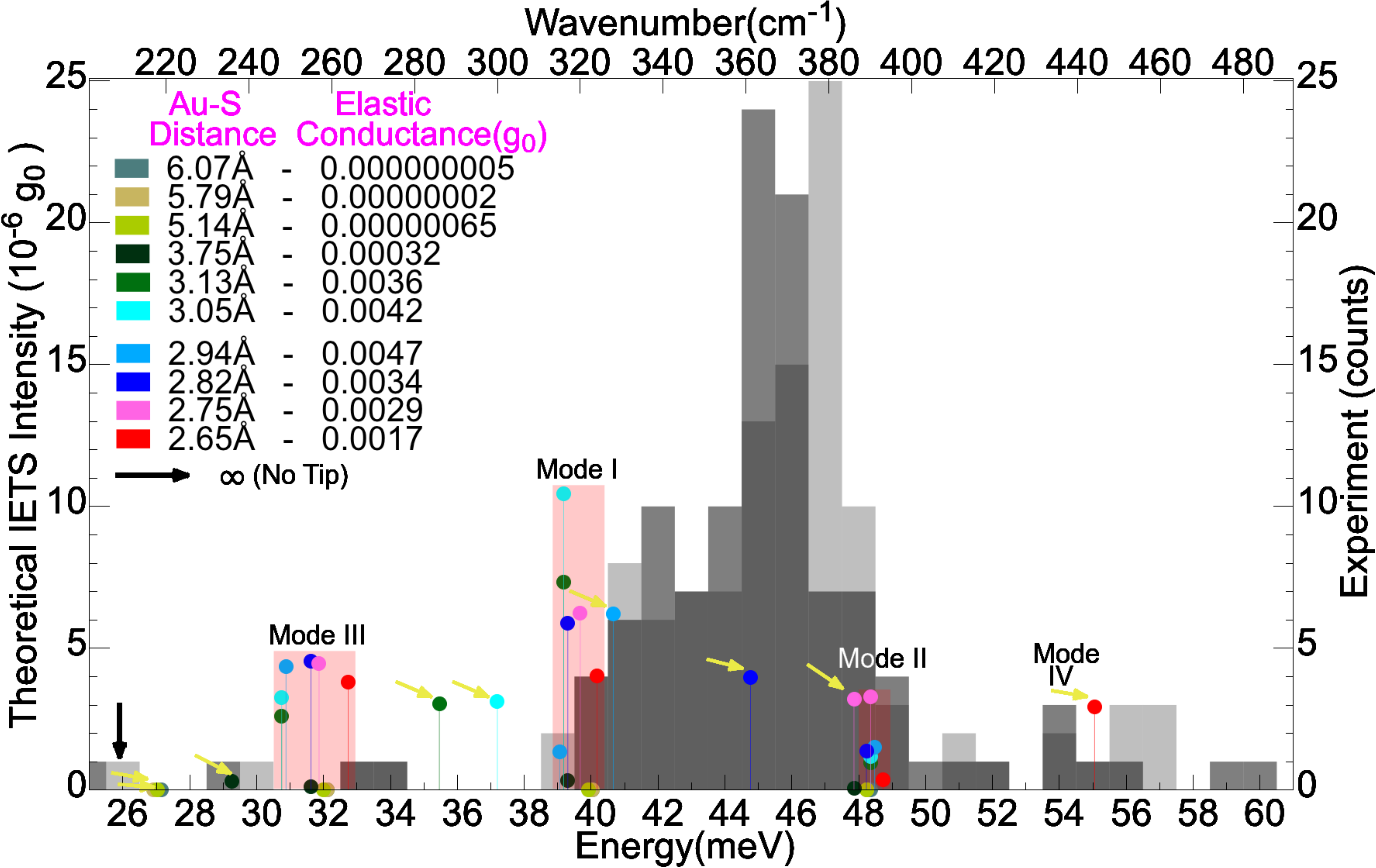}     }
\subfigure[] {   \label{Fig4(c)}
\includegraphics[width=0.485\linewidth]{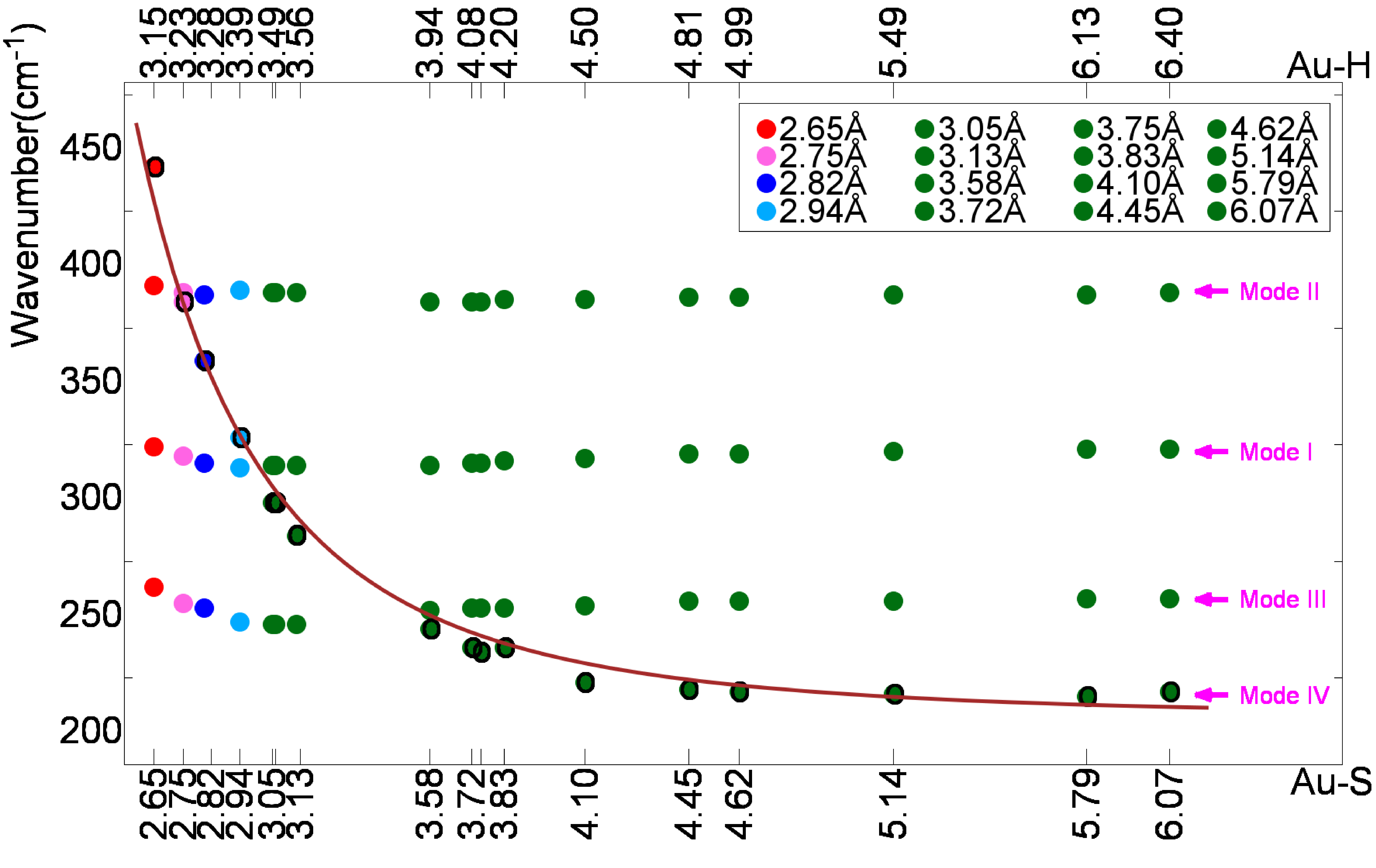}     }
\caption{{\color{red}{(Color online.) }}
\subref{Fig4(a)} I to IV: Calculated geometries of PDT molecular junctions with a H atom bound to one S atom of the molecule. C, H, S and Au atoms are black, grey, yellow and amber, respectively. In the relaxations the separation of the atoms of the two Au clusters furthest from the molecule was held fixed. Au--S distances are shown. V: Molecular junction relaxed without constraints and with no H atom attached to either S atom.
\subref{Fig4(b)} 
Theoretical inelastic tunneling spectra of the PDT wires, some of which are shown in \subref{Fig4(a)}, color coded according to the Au--S distance for the S atom with an attached H atom. The calculated low bias elastic conductance of each structure is shown in the legend. Chartreuse arrows indicate mode IV. Black arrow indicates the energy of mode IV if there is no Au cluster near the S--H group. Pink areas indicate energy ranges in which modes I, II and III occur.  The experimental phonon mode histogram (grey) of Hihath \emph{et al.}\cite {HihathArroyoRubio-BollingerTao08} is shown as a reference.
\subref{Fig4(c)} 
Au-S and thiol H vibrational modes are plotted vs. Au-S distance; Au-H distances are also shown. Mode IV points are circled in black. The solid red curve is a fit of the mode IV frequencies to an inverse cube dependence on the Au--S distance; see text. 
}
\end{figure*}
%

We assume that the molecule has bonded at one of its ends to the first gold electrode via a sulfur atom with no attached hydrogen atom and that the second electrode approaches the thiol (S--H) group at the other end of the molecule. We modeled this situation by calculating the structures of extended molecules that are relaxed subject to a geometrical constraint. The role of the electrodes is again played by 14 atom gold clusters. We carried out the density functional theory-based relaxations starting from initial geometries in which the first gold cluster was located a typical bonding distance (about  2.4\AA) from the sulfur atom of the molecule with no attached hydrogen atom, while   the second gold cluster was located at a larger distance from the molecule's S--H group. During the relaxation process the distance $l$ between two gold atoms (the furthest gold atom from the molecule in each of the two gold clusters) was kept fixed. By choosing different values of $l$ we obtained relaxed extended molecule geometries with differing values of the distance between the molecule's  S--H group and the nearest gold cluster. Some representative examples of relaxed geometries obtained in this way as the gold cluster representing the STM tip approaches the molecule's thiol group are shown in  Fig.~\ref{Fig4(a)} I - IV, where the gold-sulfur distances in each junction are also shown. A similar junction but with no thiol hydrogen atoms and relaxed without geometrical constraints is shown in Fig.~\ref{Fig4(a)} V for comparison. Notice how the orientation of the PDT molecule changes from Fig.~\ref{Fig4(a)} I to Fig.~\ref{Fig4(a)} IV so as to assume an energetically favorable bonding configuration in Fig.~\ref{Fig4(a)} where the gold-sulfur bond has fully formed.

The calculated inelastic tunneling spectra for a series of structures obtained in this way are shown in Fig.~\ref{Fig4(b)}. 
The S--Au distance for the sulfur atom with the attached hydrogen atom for each structure is shown in the
legend together with the calculated low bias elastic conductance of the junction and the color used to represent the inelastic tunneling spectrum for that structure.
 The experimental IETS histogram of Hihath \emph{et al.}\cite{HihathArroyoRubio-BollingerTao08} is also included in Fig.~\ref{Fig4(b)} but in this figure it serves as a reference only since, as we have discussed in Section \ref{IETSresults}, it is plausible that most of the counts in the histogram were obtained from junctions with no hydrogen atom attached to either sulfur atom of the molecule being probed. 
 The shaded pink areas indicate the energy ranges in which we find modes I, II and III (as defined in Fig. \ref{Fig2(a)}) to occur for these molecular junctions. 

The most striking feature of Fig.~\ref{Fig4(b)} is the behavior of mode IV that (as in Fig. \ref{Fig2(a)}) is characterized by a strong vibrational amplitude of the thiol hydrogen atom, and is marked by the chartreuse colored arrows in Fig.~\ref{Fig4(b)}. The energy of this mode is $\sim 26$ meV (marked by the black arrow) when there is no gold cluster in the vicinity of the S--H group, i.e., the STM tip is at infinity.  
As the gold cluster approaches the S--H group the mode IV energy increases dramatically, reaching a value of $\sim 55$ meV when the gold-sulfur distance reaches the value of 2.65\AA\ at which the chemical bond between the gold and sulfur atoms in the presence of the thiol hydrogen atom has fully formed. This doubling of the mode IV energy as the gold STM tip approaches the molecule is a {\em large} effect and therefore by tracking the energy of mode IV it should be possible to monitor the approach of an STM tip to a molecule adsorbed on a surface experimentally. The disappearance of mode IV from the inelastic tunneling spectrum and the simultaneous jump in the energy of mode I from $\sim 40$ meV to $\sim 46$ would then signal the ultimate detachment of the thiol hydrogen atom from the sulfur atom in such an experiment. 

In Fig.~\ref{Fig4(c)} we plot the wavenumbers of modes I - IV vs the Au -- S distance as the distance between the 
gold cluster and the molecule is varied. The mode IV results are circled in black for clarity. 
While the energies of all of the vibrational modes plotted show some variation with the Au -- S distance, that of mode IV is by far the largest. As can be seen in Fig.~\ref{Fig4(c)}, this variation of mode IV fits well to the empirical form $n = 208 + 486(x-1.35)^{-3}$, the red curve in Fig.~\ref{Fig4(c)}.  Here $n$ is the mode IV wavenumber in cm$^{-1}$ and $x$ is the Au -- S distance in Angstoms.  A similar (bond length to the negative third power) dependence of molecular vibrational frequencies on bond length was proposed empirically by Philip M. Morse in 1929 \cite{Morse29} and later discussed by other authors \cite{Gans71}. We note, however, that there are significant differences between the systems considered here and those in the early work  \cite{Morse29,Gans71} on the dependence of vibrational frequencies and the associated force constants on bond length: The vibrational modes considered in the previous work\cite{Morse29,Gans71} were stretching modes, while mode IV considered here is torsional. Also the bond lengths considered previously\cite{Morse29,Gans71} were equilibrium bond lengths that differed because the bonds were in different molecules, whereas in the present work we are considering non-equilibrium interatomic distances that differ due to differences in the geometrical constraints applied to the system.


\section{Conclusions} \label{Conclusions}

We have investigated theoretically the conformations, the elastic low bias conductances and the inelastic tunneling spectra of molecular junctions containing propanedithiol and propanedithiolate molecules bridging gold electrodes. We found that, for junctions in which the molecule bonds to a gold electrode via a sulfur atom that retains its hydrogen atom, a relaxed structure of the junction could be obtained provided that the sulfur atom bonds to a single gold atom of the electrode. This differs from propanedithiolate molecules bonding to gold via a sulfur atoms that have lost their thiol hydrogen atoms in which case relaxed geometries with a sulfur atom bonded to two or three gold atoms have also been obtained\cite{Demir2011, Demir2012}. We have demonstrated that inelastic tunneling spectroscopy  should be able to distinguish between junctions in which neither sulfur atom retains its thiol hydrogen atom, one of the two sulfur atoms retains its thiol hydrogen atom, or both sulfur atoms retain their thiol hydrogen atoms. 

Comparison between our results for junctions relaxed with no geometrical constraints and the experimental inelastic tunneling data of Hihath \emph{et al.} \cite{HihathArroyoRubio-BollingerTao08} indicates that most of their experimental measurements were carried out on junctions in which neither sulfur atom of the PDT molecule carrying the current retained its thiol hydrogen atom. We found no evidence of features in their experimental inelastic tunneling histogram\cite{HihathArroyoRubio-BollingerTao08} that might signal conduction via a propanedithiol molecule retaining both of its thiol hydrogen atoms. However, some weak features in the experimental inelastic tunneling histogram\cite{HihathArroyoRubio-BollingerTao08} are consistent with the possibility of a small fraction of their junctions being bridged by molecules with one sulfur atom retaining its hydrogen atom, although molecules with gauche geometries and no thiol hydrogen atoms may also be responsible for these spectral features. Further experimental studies with greater sensitivity to weak features of the inelastic tunneling spectra are needed to distinguish between these possibilities.   

We also studied the evolution of the inelastic tunneling spectrum as a gold STM tip approaches an intact S--H group at one end of a propanedithiol(ate) molecule that is bonded to a second gold electrode via a sulfur atom that has lost its thiol hydrogen atom. We found that the frequency of a vibrational mode associated with torsional motion of the thiol hydrogen atom increases by approximately a factor of two as the STM tip approaches the molecule. Since we predict this mode to have a moderately strong intensity in the inelastic tunneling spectrum this frequency shift can be used to monitor the separation between the gold STM tip and the thiol group experimentally as the tip approaches the molecule. When the thiol hydrogen atom detaches from the molecule we predict this vibrational mode to disappear from the inelastic tunneling spectrum and the energy of the strongest IETS feature associated with vibrations of the sulfur to simultaneously increase by $\sim$4-5meV while its intensity also increases. 

Experiments observing this predicted behavior would be of interest since they would shed much needed light on how and when the thiol hydrogen atom detaches from the sulfur atom during bond formation between molecules and gold electrodes in single-molecular junctions. They may also help clarify the reasons why apparently similar experiments on gold-thiol molecular junctions often yield measured conductances differing by orders of magnitude \cite{conductancevalues}. 

\section{Acknowledgments}
This research was supported by CIFAR, NSERC, Westgrid, Compute Canada and Sharcnet. 
We thank J. Hihath, N. J. Tao, E. Emberly, N. R. Branda, V. E. Williams and A. Saffarzadeh 
for helpful discussions and J. Hihath and N. J. Tao for providing to us their experimental data 
in digital form.


\end{document}